\documentclass[12pt,preprint]{aastex}
\begin{document}
\newcommand{\calu}{{\cal U}}
\newcommand{\calq}{{\cal Q}}
\newcommand{\bx}{{\rm \bf x}}
\newcommand{\bk}{{\bar{\kappa}}}
\title{The $\Omega_{DE}-\Omega_{M}$ Plane in Dark Energy Cosmology}
\author{Yuan Qiang$^{1}$, Tong-Jie Zhang$^{1,2}$}
\affil{$^1$Department of Astronomy, Beijing Normal University,
Beijing 100875, P.R.China; tjzhang@bnu.edu.cn \\
$^2$Canadian Institute for Theoretical Astrophysics(CITA),
University of Toronto,Toronto,ON M5S3H8,Canada;
tzhang@cita.utoronto.ca}

\begin{abstract}
The dark energy cosmology with the equation of state $w=constant$
is considered in this paper. The $\Omega_{DE}-\Omega_{M}$ plane
has been used to study the present state and expansion history of
the universe. Through the mathematical analysis, we give the
theoretical constraint of cosmological parameters. Together with
some observations such as the transition redshift from
deceleration to acceleration, more precise constraint on
cosmological parameters can be acquired.
\end{abstract}


\maketitle
\section{Introduction}

One of the greatest challenges in modern cosmology is
understanding the nature of the observed evolution status of the
universe in late-time phase. The type Ia supernovae (SNe Ia)
searches \citep{1998AJ....116.1009R,1999ApJ...517..565P}, the
cosmic microwave background (CMB) results from balloon and ground
experiments
\citep{1999ApJ...524L...1M,2000Natur.404..955D,2000ApJ...545L...5H,
2002ApJ...568...38H,2003ApJ...591..540M,2003A&A...399L..25B} and
recent WMAP \citep{2003ApJS..148..175S} and recent WMAP
\citep{2003ApJS..148..175S} observation all suggest that the
universe is spatially flat and undergoing a phase of accelerating
at the present time due to the current domination of some sort of
negative-pressure dark energy (DE). The dark energy is usually
characterized by a parameter of an equation-of-state (hereafter,
EOS) $w\equiv p/\rho $, the ratio of the spatially-homogeneous
dark-energy pressure $p$ to its energy density $\rho $. The
cosmological constant
\citep{1989RvMP...61....1W,1992ARA&A..30..499C,
1995Natur.377..600O} of order (10$^{-3}$eV)$^{4}$, the EOS of
which $w=-1$, is the simplest candidate for dark energy. However,
it is 120 orders of magnitude smaller than the naive expectations
from quantum field theory. Another widely explored possibility is
quintessence \citep{1988PhRvD..37.3406R,1997PhRvD..55.1851C,
1998PhRvL..80.1582C}, which is described in terms of a cosmic
scalar field $\phi $. Thus in such models, the EOS takes
$-1<w<-1/3$, and the dark-energy density decreases with scale
factor $a(t)$ as $\rho \propto a^{-3(1+w)}$.

The $\Omega_{DE}-\Omega_{M}$ plane is one of the most fundamental
diagrams in modern observational cosmology, which has been used to
study the present state and expansion history of the universe.  To
study the cosmological dynamics of the universe for different
scenarios the phase plane analysis is used. In this paper, we
focus on the case of EOS $w=constant$ for simplicity. Three
choices of $w$ are taken to be examples of our analysis, however,
this analysis method can be appropriate for arbitrary $w$. The
case of $w=-1$, which represents the dark energy is a constant
independent of cosmic time (the cosmological constant), is the
simplest choice for dark energy. Furthermore, it was strongly
supported by SNe Ia \citep{2004ApJ...607..665R} and CMB
\citep{2003ApJS..148..175S} observations. Another candidate for
dark energy is the cosmic topological defect
\citep{2003RvMP...75..559P} such as cosmic string
\citep{1994csot.book.....V} and domain wall
\citep{1999astro.ph..8047B}, and so on. The EOS of topological
defect is $w=-n/3$, where $n$ is the dimension of defect. $n=1$
(accordingly $w=-1/3$) corresponds to cosmic string and $n=2$
($w=-2/3$) corresponds to domain wall respectively. In addition,
these choices of $w$ are easy to be calculated analytically.

This paper is organized as follows: in Sec.~\ref{eos} we discuss
the restrictions which can be set on this cosmological model from
the astronomical observations, and in Sec.~\ref{plane} the
$\Omega_{DE} -\Omega_M$ plane of the three cases ($w=-1,\,-1/3$
and $-2/3$) is discussed in detail. We study the transit redshift
from deceleration to acceleration in the $\Omega_{DE} -\Omega_M$
plane in Sec.~\ref{redshift}. The discussion of the results and
their further possible generalization is presented in
Sec.~\ref{cd}.

\section{The Equation of State for Dark Energy}

\label{eos} For most purpose, we consider a general dark energy
EOS $w(z)$ which varies with the cosmic time $t$ or redshift $z$
\begin{equation}  \label{H^2}
w(z)=\frac{p_{DE}(z)}{\rho_{DE}(z)},
\end{equation}
where $p_{DE}(z)$ and $\rho_{DE}(z)$ are the time-dependent pressure and
energy density respectively and redshift $z$ is defined by scale factor $a$,
$1+z=a_0/a$. Using the conservation of energy, we can express the energy
density of dark energy by
\begin{equation}
f_{DE}(z)=\frac{\rho_{DE}(z)}{\rho_{DE0}}=\exp\,[\int_{0}^{z}3(1+w_{DE}(z))
\mathrm{d}\ln (1+z)].  \label{rhox}
\end{equation}
Thus Friedmann equation neglecting cosmic radiation can be expressed as
\begin{equation}
E^{2}(z)=\frac{H^{2}(z)}{H_{0}^{2}} =\Omega_{M}(1+z)^{3}+
\Omega_{DE}f_{DE}(z)+\Omega_{k}(1+z)^{2},  \label{E2z}
\end{equation}
where $\Omega_{M}$, $\Omega_{DE}$ and $\Omega_{k}$ are
respectively density parameters at the present epoch $t_0$. More
generally, we can rewrite Eq.(\ref{E2z}) in terms of cosmological
density parameters at redshift $z$
\begin{equation}
\Omega^z_M+\Omega^z_{DE}+\Omega^z_{k}=1,  \label{Omegazp1}
\end{equation}
where
\begin{equation}
\left \{
\begin{array}{ll}
\Omega^z_M=\frac{\rho_{M}}{\rho_{c}}=\frac{\rho_{M0}(1+z)^{3}} {
\rho_{c0}E^{2}(z)}=\frac{\Omega_{M}(1+z)^{3}}{E^{2}(z)} &  \\
\Omega^z_{DE}=\frac{\rho_{DE}}{\rho_{c}}=\frac{\rho_{DE0}f_{DE}} {
\rho_{c0}E^{2}(z)}=\frac{\Omega_{DE}f_{DE}(z)}{E^{2}(z)} &  \\
\Omega^z_k=-\frac{kc^{2}}{a^{2}(z)H^{2}(z)}=
\frac{\Omega_{k}(1+z)^2}{E(z)^2}  &
\end{array}
\right. , \label{Omegaz3}
\end{equation}
and the critical density at redshift $z$ is
\begin{equation}
\rho_{c}=\frac{3H^{2}}{8\pi G}=\frac{3H^{2}_{0}}{8\pi G}E^{2}(z)=%
\rho_{c0}E^{2}(z).
\end{equation}

\begin{figure}[t]
\includegraphics[width=0.8\textwidth]{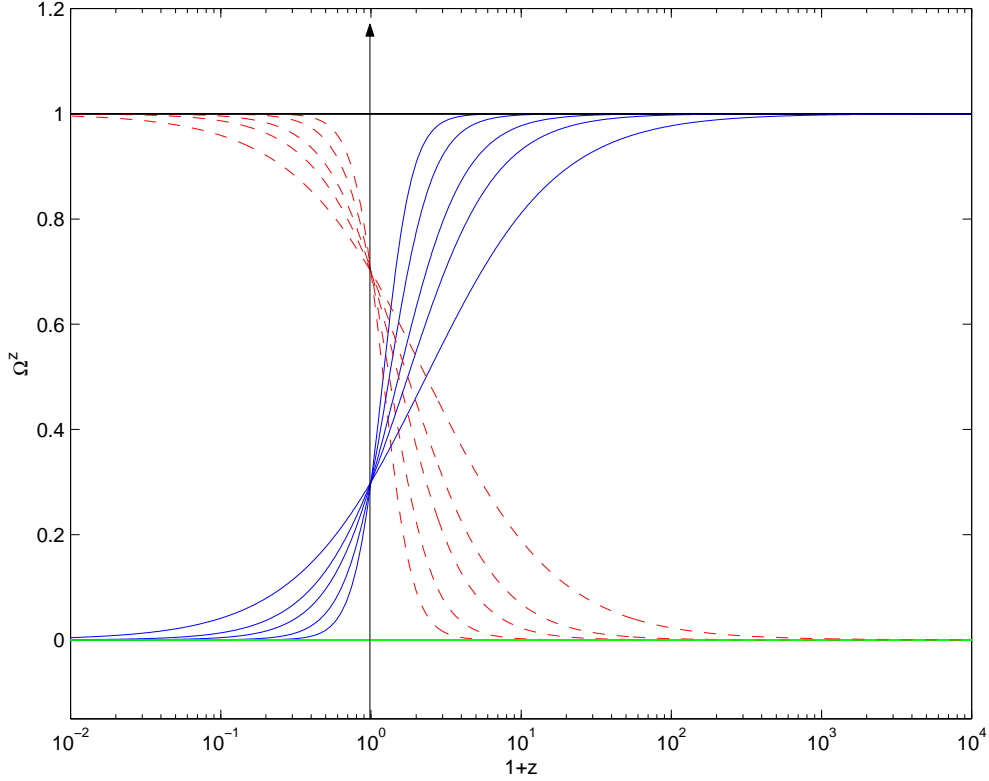}
\caption{The evolution of cosmological density parameters $
\Omega^z_{M,\,DE,\,k,\,T}$ as a function of redshift $z$, where we
have assumed a spatially flat concordant cosmological model:
$\Omega_M=0.3$ and $\Omega_{DE}=0.7$ with $w=-1 $ and
$\Omega_k=0$. The solid and dashed lines denote the evolution of
$\Omega^z_M$ and $\Omega^z_{DE}$, while the two horizonal straight
lines correspond to that of $\Omega^z_T$ and $\Omega^z_k$ ,
respectively. From left to right for $z>0$ and from right to left
for $z<0$, the solid lines representing $\Omega^z_M$ and the
dashed lines representing $\Omega^z_{DE}$ correspond to the cases
of $w =-3/2,-1,-2/3,-1/2$ and $-1/3$.} \label{fig1eps}
\end{figure}

If $w_{DE}(z)=w$ (constant) is independent of time, the expression for $\rho
_{DE}(z)$ above becomes
\begin{equation}
f_{DE}(z)=\frac{\rho _{DE}(z)}{\rho _{DE0}}=(1+z)^{3(1+w)}\propto
a^{-3(1+w)},
\end{equation}
and the expression of $E(z)$ reduces to
\begin{equation}
E(z)=\sqrt{\Omega _{M}(1+z)^{3}+\Omega _{DE}(1+z)^{3(1+w)}+\Omega
_{k}(1+z)^{2}}.
\end{equation}
Throughout this paper, we just consider constant EOS $w$ rather
the variation of its $w(t)$. At the present epoch,
Eq.(\ref{Omegazp1}) satisfies $\Omega _{M}+\Omega _{DE}+\Omega
_{k}=1$. For spatially flat universe, $ \Omega _{k}=0$, i.e.
$\Omega _{M}+\Omega _{DE}=1$, then
\begin{equation}
\Omega _{T}^{z}=\Omega _{M}^{z}+\Omega _{DE}^{z}=\frac{\Omega
_{M}(1+z)^{3}}{ E^{2}(z)}+\frac{\Omega
_{DE}(1+z)^{3(1+w)}}{E^{2}(z)}=1,
\end{equation}
which means $\Omega _{T}^{z}$ will equal unit at any resdshift $z$
if only the universe is spatially flat today. Fig.\ref{fig1eps}
draw the cosmological density parameters $\Omega ^{z}$ as a
function of redshift $z$ in various spatially flat cosmologies.
With the expansion of the universe from the Big Bang (redshift
$z=\infty $) to the future ($z=-1$) theoretically, matter density
$\Omega _{M}^{z}$ drops to zero from unit, while dark energy
density $\Omega _{DE}^{z}$ reaches unit from zero. The universe
start from Einstein-de Sitter model and end in a de-Sitter phase.
This point can also be demonstrated theoretically by taking the
limit of Eq.(\ref{Omegaz3})
\begin{eqnarray}
& &\lim_{z \to \infty} \Omega _{M}^{z}=\lim_{z\to \infty}
\frac{\Omega _{M}(1+z)^{3}}{E^{2}(z)}=1,\ \lim_{z\to -1}
\Omega _{M}^{z}=0;  \nonumber\\
& &\lim_{z\to \infty}\Omega _{DE}^{z}=\lim_{z\to \infty}
\frac{\Omega _{DE}\,\exp [3\int_{0}^{z}(1+w(z))\mathrm{d}\,\ln
(1+z)]} {E^{2}(z)}=0,\ \lim_{z\to -1}\Omega _{DE}^{z}=1.
\label{limtOmega}
\end{eqnarray}

\section{The $\Omega_{DE}-\Omega_{M}$ Plane and the Expansion History
of the Universe}

\label{plane} Using Eq.(\ref{Omegaz3}), we can relate the density
parameters $\Omega^z_{M,\,DE,\,k}$ at $z$ and their current ones
$\Omega_{M,\,DE,\,k}$ by
\begin{equation}
\frac{(\Omega^z_M+\Omega^z_{DE}-1)^3}{(\Omega^z_M)^{(1+3w)/w}(%
\Omega^z_{DE})^{-1/w}}= \frac{(\Omega_M+\Omega_{DE}-1)^3}{%
\Omega^{(1+3w)/w}_M\Omega^{-1/w}_{DE}}=C.  \label{omegaplanew}
\end{equation}
Given the value of the constant $C$, one can represent the specific
cosmological models by the second term and the expansion history of the
universe by the first term in Eq.(\ref{omegaplanew}) in the $%
\Omega_{DE}-\Omega_{M}$ plane respectively. This leads to the
identity of the the $\Omega_{DE}-\Omega_{M}$ plane and the
expansion history of the universe in the $\Omega_{DE}-\Omega_{M}$
plane.

The universe can be roughly divided into three types: open, flat
and closed. In the absence of the exotic dark energy, the fate of
the universe are easily to be understood. At the presence of dark
energy, the situation becomes more complicated. Especially, the
recollapse of the universe can exist in some types of cosmological
models. From the definition of expansion rate of the universe,
Eq.(\ref{E2z}), one can guarantee the existence of recollapse by
the criterion $\dot{a}(t)=0$ at some moment $t$
\begin{equation}
\frac{H^2(z)}{H_{0}^2}=\frac{(\dot{a}/a)^2}{H_{0}^2}=E^2(z)=
\Omega_{M}(1+z)^{3}+\Omega_{DE}f_{DE}(z)+\Omega_{k}(1+z)^{2}=0.  \label{E2z0}
\end{equation}
Setting $x=a/a_0=1/(1+z)$, we can see that when $z$ varies from
$+\infty$ to $-1$, $x$ varies from 0 to $+\infty$. Defining a
function
\begin{equation}
F(x)=\Omega_{DE}x^{-3w}+\Omega_{k}x+\Omega_{M},  \label{Fx}
\end{equation}
we can easily find that $E^2(z)\geq0\Leftrightarrow F(x)\geq0$,
and if only there exists a positive root to $F(x)$, the universe
can recollapse at some time $a(t)$.

\subsection{The Case of $w=-1$}

For the case of cosmological constant, $w=-1$, we have
\begin{equation}
\frac{(\Omega^z_M+\Omega^z_{DE}-1)^3}{(\Omega^z_M)^2\Omega^z_{DE}}= \frac{%
(\Omega_M+\Omega_{DE}-1)^3}{\Omega^2_M\Omega_{DE}}=C,  \label{omegaplane1}
\end{equation}
for $\Omega_{DE}-\Omega_{M}$ plane and the expansion trajectory of
the universe in the $\Omega_{DE}-\Omega_{M}$ plane, and
\begin{equation}
F(x)=\Omega_{DE}x^3+\Omega_{k}x+\Omega_{M}.  \label{Fx1}
\end{equation}
Given the value of $C$, one can follow the expansion trajectory of
the universe in terms of the density parameters $\Omega^z$ at any
redshift $z$, which, essentially, is identical to the $
\Omega_{DE}-\Omega_{M}$ plane that represents the location of
specific cosmological models in terms of current density
parameters $\Omega$. Parameter $C$ cannot be arbitrary for the
constrain condition $F(x)\geq 0$ (otherwise, $E^2(z)=H^2/H^2_0<0$,
it is impossible). So we get $\Omega_{DE}\geq0$. Below we just
consider the situation of $\Omega_{DE}>0$, the case of
$\Omega_{DE}=0$ is just the limit of $C\rightarrow-\infty$(see
below). Taking the derivation of $F(x)$, we have
\begin{equation}
F^{\prime}(x)=3\Omega_{DE}x^2+\Omega_{k}. \label{Fx1p}
\end{equation}

For the cases $\Omega_k>0\,(k=-1)$ and $\Omega_k=0\,(k=0)$, which
correspond to $C<0$ and $C=0$ respectively, $F^{\prime}(x)>0$, so
$F(x)> F(0)=\Omega_M>0$ will be satisfied for all values of $x$.
There is something different for the case $\Omega_k<0\,(k=1)$
which corresponds to $C>0$. There is a positive root
$x=x_0=\sqrt{-\Omega_k/3\Omega_{DE}}$ for equation
$F^{\prime}(x)=0$. And we can further get that $x_0$ is the
minimum point of $F(x)$ for $F^{\prime\prime}(x_0)>0$. If only
$F(x_0)\geq0$, $F(x)\geq0$. So this condition leads to
$C\leq27/4$. To sum up, the value of $C$ can't be arbitrary. There
is a basic constraint from the standard cosmology. For the case of
$w=-1$, $C$ has to be in the range (-$\infty,27/4$]. Only when
$C=27/4$, $F(x_0)=0$, that is to say $F(x)$ has a positive root
$x=x_0$, the universe can recollapse at some time; when $C<27/4$,
the universe will expand forever; and when $C>27/4$, the universe
is impossible to exist as a physical one, or to say it is not a
Big Bang universe.

\begin{figure}[t]
\includegraphics[width=0.8\textwidth]{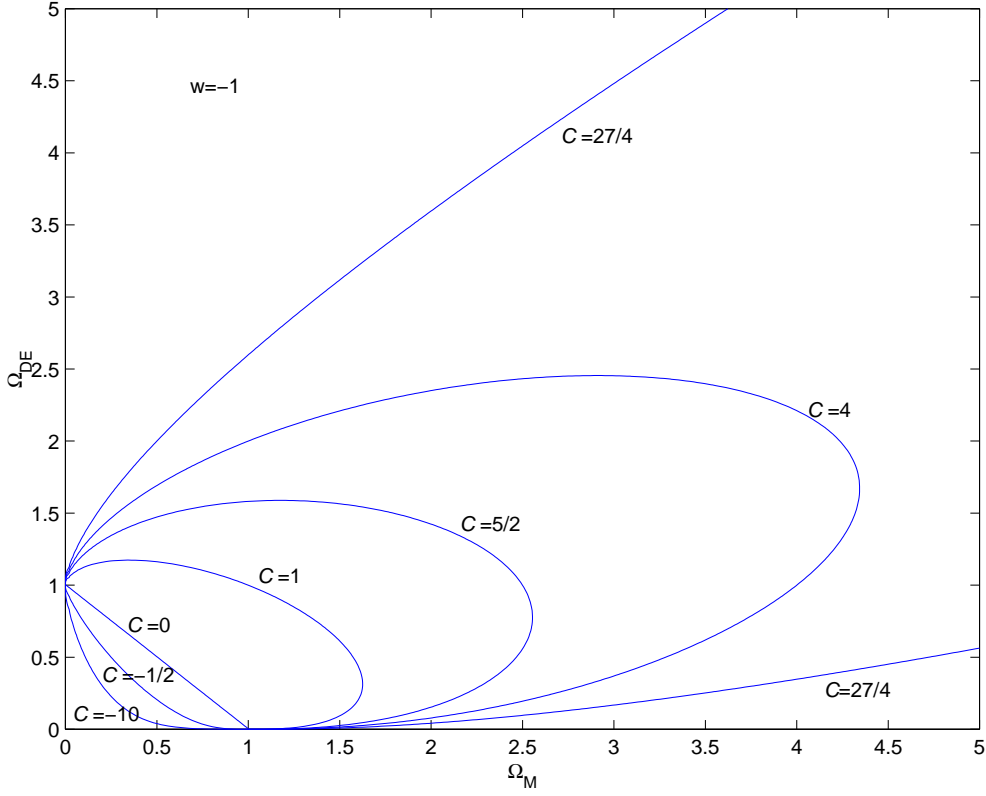}
\caption{The $\Omega_{DE}-\Omega_{M}$ plane for the case of $w=-1$
(cosmological constant). The permitted region is the interior of
the line labelled $C=27/4$.} \label{fig2eps}
\end{figure}

These results are displayed in Fig.\ref{fig2eps}. The permitted
region is the interior of the line labelled $C=27/4$.

\subsection{The Case of $w=-1/3$}

In this case, Eq.(\ref{omegaplanew}) describing the $\Omega_{DE}-
\Omega_{M}$ plane and the expansion trajectory of the universe
reduces to
\begin{equation}
\frac{\Omega^z_M+\Omega^z_{DE}-1}{\Omega^z_{DE}}= \frac{\Omega_M+
\Omega_{DE}-1}{\Omega_{DE}}=\sqrt[3]{C},  \label{omegaplane13}
\end{equation}
and the function $F(x)$ becomes
\begin{equation}
F(x)=(\Omega_{DE}+\Omega_{k})x+\Omega_{M},  \label{Fx2}
\end{equation}
which can be further written as
\begin{equation}
F(x)=(1-\Omega_M)x+\Omega_{M}.  \label{Fx22}
\end{equation}

From the condition $F(x)\geq0$, we get $0<\Omega_M\leq 1$ (it can
be easily understood that $\Omega_M>0$ originates from the fact
that the matter density can't be negative). According to
Eq.(\ref{omegaplane13}), we have
$0<\Omega_M=(\sqrt[3]{C}-1)\Omega_{DE}+1\leq1$, so
\begin{equation}
\left\{
\begin{array}{ll}
-\infty<C\leq1,\ \ &\ \ \rm{for}\ \Omega_{DE}>0\\
C=\pm\infty,\ \ &\ \ \rm{for}\ \Omega_{DE}=0\\
1\leq C<+\infty,\ \ &\ \ \rm{for}\ \Omega_{DE}<0\\
\end{array}
\right..   \label{OmegaDE}
\end{equation}

\begin{figure}[t]
\includegraphics[width=0.8\textwidth]{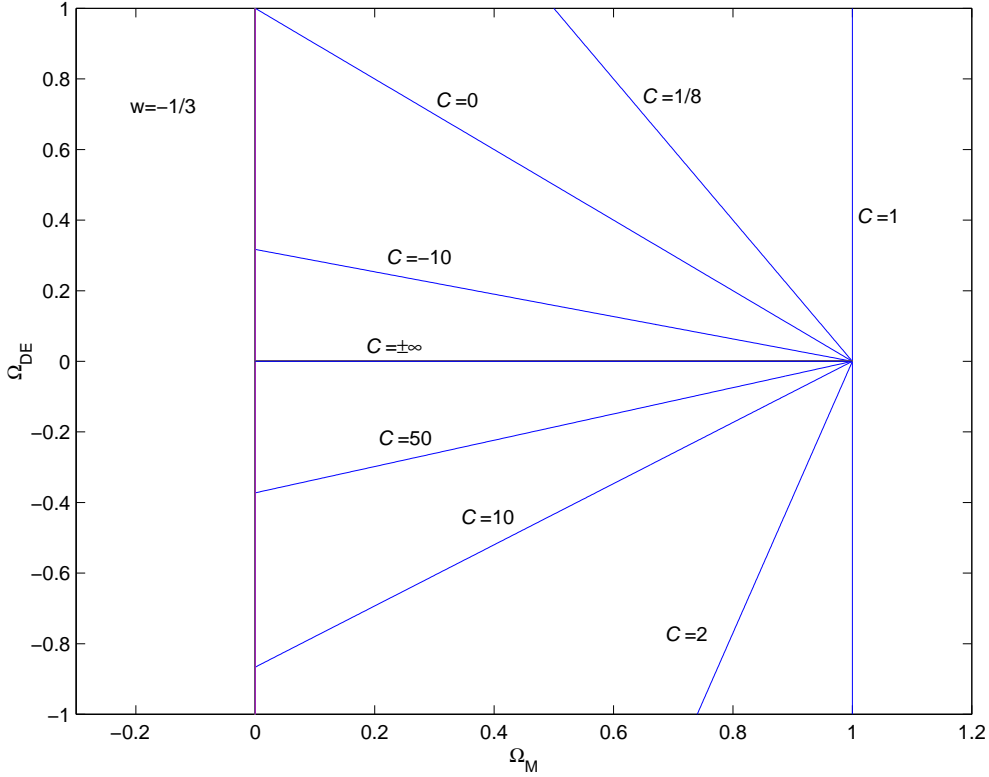}
\caption{The same as Fig.\protect\ref{fig2eps} but for the case of
$w=-1/3$. The permitted region of $\Omega_{DE}$ and $\Omega_M$ is
between the two vertical line.} \label{fig3eps}
\end{figure}

These results are displayed in Fig.\ref{fig3eps}. And there isn't
a positive root for function $F(x)$, so this kind of universe will
expand forever.

\subsection{The Case of $w=-2/3$}

In this case, Eq.(\ref{omegaplanew})reduces to
\begin{equation}
\frac{\Omega^z_M+\Omega^z_{DE}-1}{\sqrt{\Omega^z_M}\sqrt{\Omega^z_{DE}}}=
\frac{\Omega_M+\Omega_{DE}-1}{\sqrt{\Omega_M}\sqrt{\Omega_{DE}}}=\sqrt[3]{C},
\label{omegaplane23}
\end{equation}
and the function $F(x)$ becomes
\begin{equation}
F(x)=\Omega_{DE}x^2+\Omega_{k}x+\Omega_{M}.  \label{Fx3}
\end{equation}
$F(x)\geq0$ needs $\Omega_{DE}\geq0$ (also we just consider
$\Omega_{DE}>0$). The same as done in the case $w=-1$, we study
the derivation of $F(x)$
\begin{equation}
F^{\prime}(x)=2\Omega_{DE}x+\Omega_{k}. \label{Fx3p}
\end{equation}

For the cases $\Omega_k>0\,(k=-1)$ and $\Omega_k=0\,(k=0)$,which
correspond to $C<0$ and $C=0$ respectively, $F^{\prime}(x)>0$, so
$F(x)> F(0)=\Omega_M>0$ will be satisfied for all values of $x$.
For the case of $\Omega_k<0\,(k=1)$ which corresponds to $C>0$,
there is a positive root $x=x_0={-\Omega_k/2\Omega_{DE}}$ of
equation $F^{\prime}(x)=0$ and $x_0$ is also the minimum point of
$F(x)$ for $F^{\prime\prime}(x_0)>0$. From $F(x_0)\geq0$ we get
$C\leq8$. So $C$ has to be in the range (-$\infty,8$]. Only when
$C=8$, $F(x)$ has a positive root $x=x_0$, the universe can
recollapse at some time; when $C<8$, the universe will expand
forever; and when $C>8$, the universe is impossible to exist as a
physical one.

\begin{figure}[t]
\includegraphics[width=0.8\textwidth]{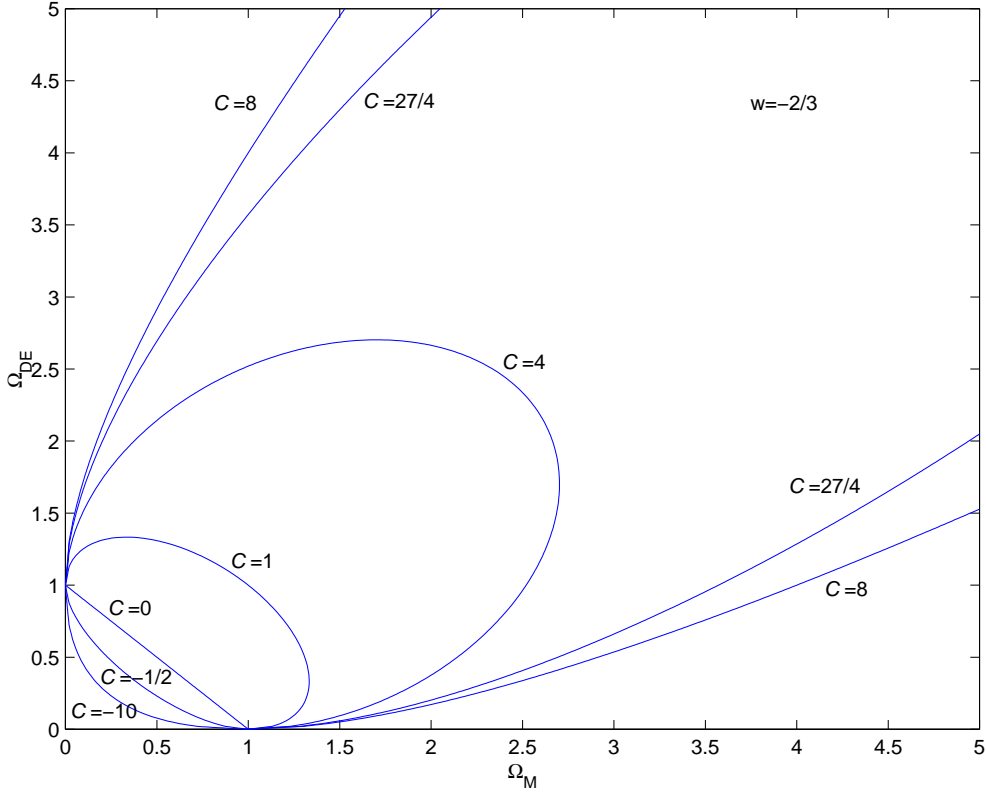}
\caption{The same as Fig.\protect\ref{fig2eps} but for the case of
$w=-2/3$. The permitted region is the interior of the line
labelled $C=8$.} \label{fig4eps}
\end{figure}

It can be shown from Fig.\ref{fig4eps} that the expansion history
of the universe are very familiar with that of cosmological
constant but with the boundary of $C=8$.

\section{The Transition Redshift from Deceleration to Acceleration in the $
\Omega_{DE}-\Omega_{M}$ Plane}

\label{redshift} 
\begin{figure}[t]
\includegraphics[width=0.8\textwidth]{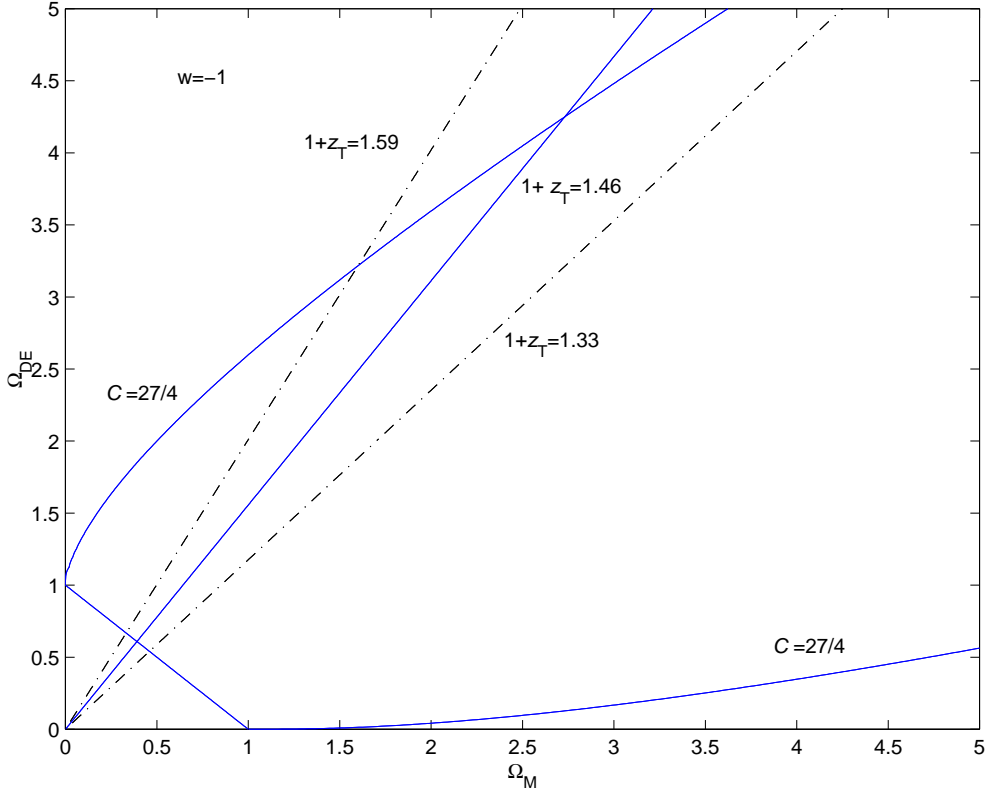}
\caption{The $\Omega_{DE}-\Omega_{M}$ plane with the best fit
estimate of transition redshift $z_T$ \citep{2004ApJ...607..665R}
for the case of $w=-1$. The curve labelled $C=27/4$ is the
boundary of this model.} \label{fig5eps}
\end{figure}

\begin{figure}[t]
\includegraphics[width=0.8\textwidth]{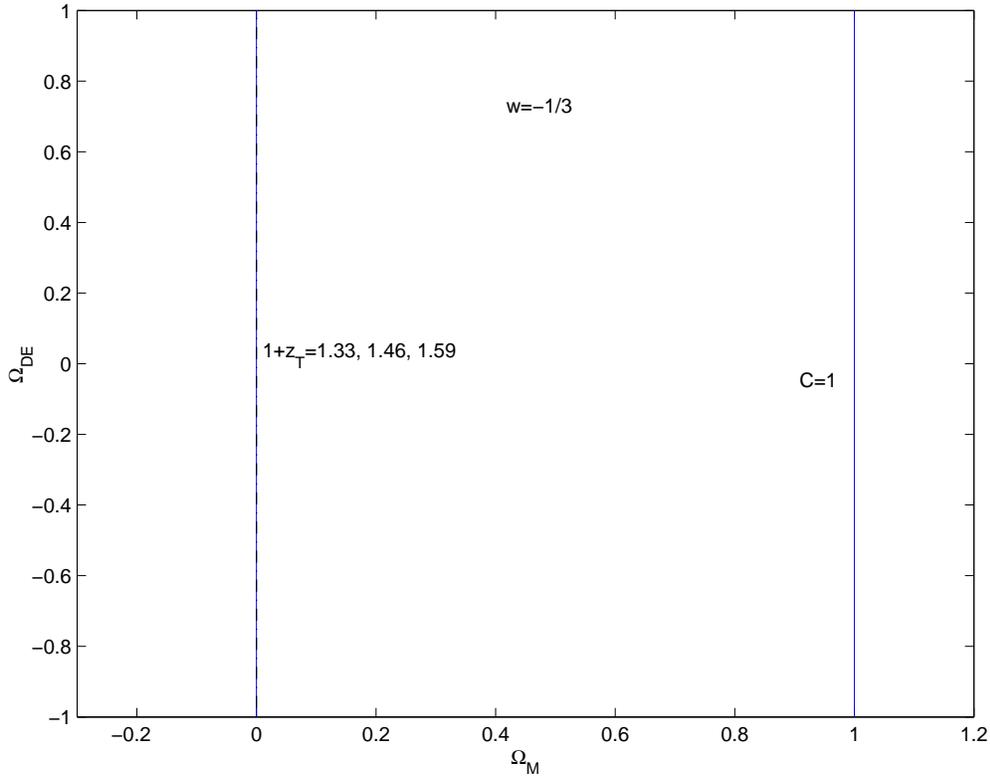}
\caption{The same as Fig.\protect\ref{fig5eps} but for the case of
$w=-1/3$. Only $\Omega_M=0$ can lead to $q(z_T)=0$.}
\label{fig6eps}
\end{figure}

\begin{figure}[t]
\includegraphics[width=0.8\textwidth]{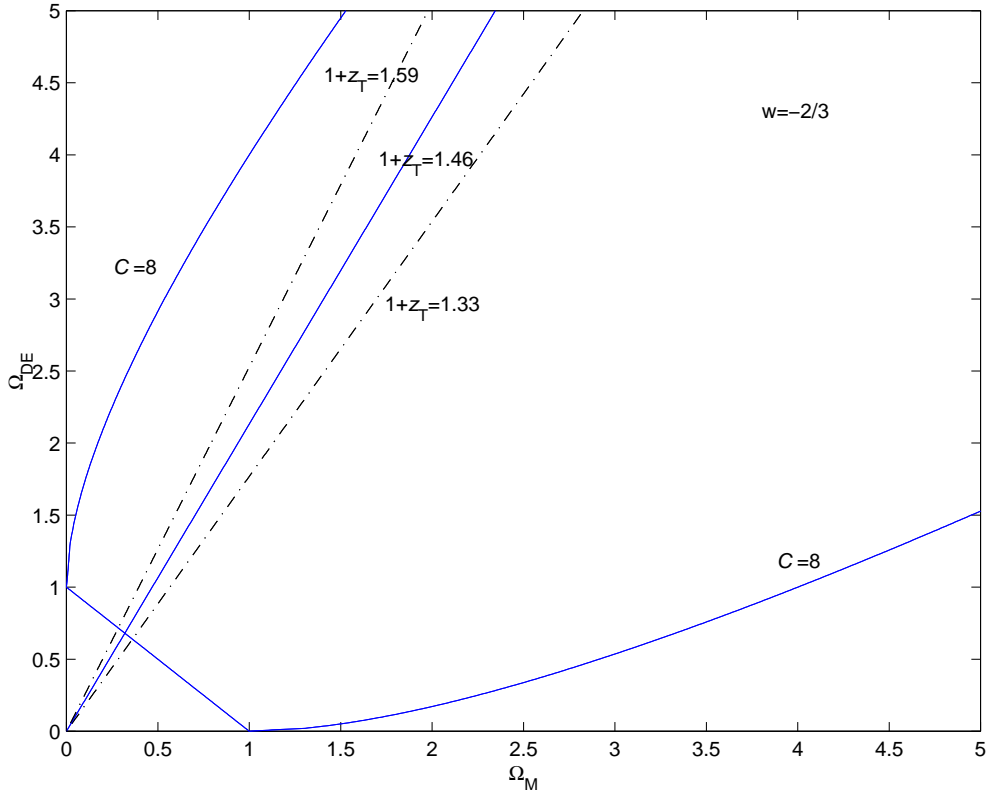}
\caption{The same as Fig.\protect\ref{fig5eps} but for the case of
$w=-2/3$. The curve labelled $C=8$ is the boundary of this model.}
\label{fig7eps}
\end{figure}

The decelerating parameter $q(z)$ is defined by
\begin{equation}
q(z)\equiv(-\frac{\ddot{a}}{a})/H^{2}(z)=\frac{1}{2E^2(z)}\frac{\mathrm{d}
E^2(z)}{\mathrm{d} z}(1+z)-1,  \label{qz}
\end{equation}
the present value $q_0=q\,(z=0)$ of which is called decelerating
factor. At the transit redshift $z_T$, the universe reaches
$\ddot{a}(z_T)=0$ or $q(z_T)=0$ and evolves from deceleration to
acceleration expansion, thus we can get the relation between
density parameters and transition redshift $z_T$ in
$\Omega_{DE}-\Omega_{M}$ plane
\begin{equation}
\Omega_{DE}=-\frac{1}{(1+3w)(1+z_T)^{3w}}\Omega_M,  \label{qzt0}
\end{equation}
which leads to
\begin{equation}
\Omega_{DE}=-\frac{\Omega_M}{(1+3w)},  \label{qzt00}
\end{equation}
at $z_T=0$ or $q_0=0$.
Clearly, as the transition redshift increases with the decrease of
$\Omega_m$. The $\Omega_{DE}-\Omega_{M}$ plane with the best fit
estimate of transition redshift $1+z_T=1.46\pm0.13$, and together
with the theoretical constraints discussed above for $w=-1,-1/3$
and $-2/3$ are shown in Fig.\ref{fig5eps}, \ref{fig6eps} and
\ref{fig7eps} respectively.




\section{Conclusion and Discussion}

\label{cd}

In this paper we examined the dynamical evolution of the universe
filled with a dark energy of EOS $w=$ constant.
Based on the Big Bang model, we give some theoretical constraints of
cosmological parameters of three special cases in the
$\Omega_{DE}-\Omega_{M}$ plane. The physical constrain condition of
$E^2(z)\geq0$ makes the cosmological parameters not be arbitrary. It
is shown in the $\Omega_{DE}-\Omega_{M}$ plane by a limited region
rather than all of the plane. Together with the observational
results of $z_T$, we can constrain the cosmological parameters more
strictly. It is shown in the Figs.\ref{fig5eps}---\ref{fig7eps} that
the observation of $z_T$ is compatible with the model of cases of
$w=-1$ and $w=-2/3$ to great extent, but is not consistent with the
one of $w=-1/3$. This has also been revealed in work on the
supernova measurements
\citep{1998ApJ...509...74G,1999ApJ...517..565P}. Apparently, the
dark energy model with $w<-1/3$ make a positive contribution to the
acceleration of the universe, while the model with $w=-1/3$ has
effect on neither the acceleration nor deceleration. In fact, the
supernova observation supports an accelerating universe. On the other
hand, because there still exists difficulty in the accurate
treatment of the behavior of cosmological defect models partly,
these models have not been very thoroughly studied
\citep{1997ApJ...491L..67S}. Thus, due to this motivation we in this
paper just make a try to test the cosmic defect models by the
properties of $\Omega_{DE}-\Omega_{M}$ plane, which maybe provides an
alternative method to test cosmological models.

In addition, when $-1<w<-1/3$, it is similar to the case of
$w=-2/3$ but is difficult to be studied analytically. From
$F(x)\geq0$ we have a general result $\Omega_{DE}\geq0$. While
$-1/3<w<0$, there should be $\Omega_k\geq0$ generally. There would
be some difficulties to get the range of $C$. For an arbitrary
value of $w$, one can give the detailed results numerically from
Eq.(\ref{omegaplanew}) and Eq.(\ref{Fx}).

\begin{acknowledgments}
We are very grateful to the anonymous referee for constructive
suggestion. This work was supported by the National Science
Foundation of China (Grants No.10473002), 985 Project, the
Scientific Research Foundation for the Returned Overseas Chinese
Scholars, State Education Ministry and the Scientific Research
Foundation for Undergraduate of Beijing Normal University (Grant
No.127002). T.J.Zhang would also like to thank Xiang-Ping Wu,
Da-Ming Chen, Bo Qin, Ue-Li Pen and Peng-Jie Zhang for their
hospitality during my visits to the cosmology groups of the
National Astronomical Observatories of P.R.China and the Canadian
Institute for Theoretical Astrophysics(CITA), University of
Toronto.
\end{acknowledgments}

\bibliography{ztjcos-lensbib}

\begin{thebibliography}{21}
\expandafter\ifx\csname natexlab\endcsname\relax\def\natexlab#1{#1}\fi

\bibitem[{{Battye} {et~al.}(1999){Battye}, {Bucher}, \&
  {Spergel}}]{1999astro.ph..8047B}
{Battye}, R.~A., {Bucher}, M., \& {Spergel}, D. 1999, ArXiv Astrophysics
  e-prints: astro-ph/9908047

\bibitem[{{Beno{\^ i}t} {et~al.}(2003){Beno{\^ i}t}, {Ade}, {Amblard},
  {Ansari}, {Aubourg}, {Bargot}, {Bartlett}, {Bernard}, {Bhatia}, {Blanchard},
  {Bock}, {Boscaleri}, {Bouchet}, {Bourrachot}, {Camus}, {Couchot}, {de
  Bernardis}, {Delabrouille}, {D{\' e}sert}, {Dor{\' e}}, {Douspis},
  {Dumoulin}, {Dupac}, {Filliatre}, {Fosalba}, {Ganga}, {Gannaway}, {Gautier},
  {Giard}, {Giraud-H{\' e}raud}, {Gispert}, {Guglielmi}, {Hamilton}, {Hanany},
  {Henrot-Versill{\' e}}, {Kaplan}, {Lagache}, {Lamarre}, {Lange},
  {Mac{\'{\i}}as-P{\' e}rez}, {Madet}, {Maffei}, {Magneville}, {Marrone},
  {Masi}, {Mayet}, {Murphy}, {Naraghi}, {Nati}, {Patanchon}, {Perrin}, {Piat},
  {Ponthieu}, {Prunet}, {Puget}, {Renault}, {Rosset}, {Santos}, {Starobinsky},
  {Strukov}, {Sudiwala}, {Teyssier}, {Tristram}, {Tucker}, {Vanel}, {Vibert},
  {Wakui}, \& {Yvon}}]{2003A&A...399L..25B}
{Beno{\^ i}t}, A., {Ade}, P., {Amblard}, A., {Ansari}, R., {Aubourg}, {\' E}.,
  {Bargot}, S., {Bartlett}, J.~G., {Bernard}, J.-P., {Bhatia}, R.~S.,
  {Blanchard}, A., {Bock}, J.~J., {Boscaleri}, A., {Bouchet}, F.~R.,
  {Bourrachot}, A., {Camus}, P., {Couchot}, F., {de Bernardis}, P.,
  {Delabrouille}, J., {D{\' e}sert}, F.-X., {Dor{\' e}}, O., {Douspis}, M.,
  {Dumoulin}, L., {Dupac}, X., {Filliatre}, P., {Fosalba}, P., {Ganga}, K.,
  {Gannaway}, F., {Gautier}, B., {Giard}, M., {Giraud-H{\' e}raud}, Y.,
  {Gispert}, R., {Guglielmi}, L., {Hamilton}, J.-C., {Hanany}, S.,
  {Henrot-Versill{\' e}}, S., {Kaplan}, J., {Lagache}, G., {Lamarre}, J.-M.,
  {Lange}, A.~E., {Mac{\'{\i}}as-P{\' e}rez}, J.~F., {Madet}, K., {Maffei}, B.,
  {Magneville}, C., {Marrone}, D.~P., {Masi}, S., {Mayet}, F., {Murphy}, A.,
  {Naraghi}, F., {Nati}, F., {Patanchon}, G., {Perrin}, G., {Piat}, M.,
  {Ponthieu}, N., {Prunet}, S., {Puget}, J.-L., {Renault}, C., {Rosset}, C.,
  {Santos}, D., {Starobinsky}, A., {Strukov}, I., {Sudiwala}, R.~V.,
  {Teyssier}, R., {Tristram}, M., {Tucker}, C., {Vanel}, J.-C., {Vibert}, D.,
  {Wakui}, E., \& {Yvon}, D. 2003, \aap, 399, L25

\bibitem[{{Caldwell} {et~al.}(1998){Caldwell}, {Dave}, \&
  {Steinhardt}}]{1998PhRvL..80.1582C}
{Caldwell}, R.~R., {Dave}, R., \& {Steinhardt}, P.~J. 1998, Physical Review
  Letters, 80, 1582

\bibitem[{{Carroll} {et~al.}(1992){Carroll}, {Press}, \&
  {Turner}}]{1992ARA&A..30..499C}
{Carroll}, S.~M., {Press}, W.~H., \& {Turner}, E.~L. 1992, \araa, 30, 499

\bibitem[{{Coble} {et~al.}(1997){Coble}, {Dodelson}, \&
  {Frieman}}]{1997PhRvD..55.1851C}
{Coble}, K., {Dodelson}, S., \& {Frieman}, J.~A. 1997, \prd, 55, 1851

\bibitem[{{de Bernardis} {et~al.}(2000){de Bernardis}, {Ade}, {Bock}, {Bond},
  {Borrill}, {Boscaleri}, {Coble}, {Crill}, {De Gasperis}, {Farese},
  {Ferreira}, {Ganga}, {Giacometti}, {Hivon}, {Hristov}, {Iacoangeli}, {Jaffe},
  {Lange}, {Martinis}, {Masi}, {Mason}, {Mauskopf}, {Melchiorri}, {Miglio},
  {Montroy}, {Netterfield}, {Pascale}, {Piacentini}, {Pogosyan}, {Prunet},
  {Rao}, {Romeo}, {Ruhl}, {Scaramuzzi}, {Sforna}, \&
  {Vittorio}}]{2000Natur.404..955D}
{de Bernardis}, P., {Ade}, P.~A.~R., {Bock}, J.~J., {Bond}, J.~R., {Borrill},
  J., {Boscaleri}, A., {Coble}, K., {Crill}, B.~P., {De Gasperis}, G.,
  {Farese}, P.~C., {Ferreira}, P.~G., {Ganga}, K., {Giacometti}, M., {Hivon},
  E., {Hristov}, V.~V., {Iacoangeli}, A., {Jaffe}, A.~H., {Lange}, A.~E.,
  {Martinis}, L., {Masi}, S., {Mason}, P.~V., {Mauskopf}, P.~D., {Melchiorri},
  A., {Miglio}, L., {Montroy}, T., {Netterfield}, C.~B., {Pascale}, E.,
  {Piacentini}, F., {Pogosyan}, D., {Prunet}, S., {Rao}, S., {Romeo}, G.,
  {Ruhl}, J.~E., {Scaramuzzi}, F., {Sforna}, D., \& {Vittorio}, N. 2000, \nat,
  404, 955

\bibitem[{{Garnavich} {et~al.}(1998){Garnavich}, {Jha}, {Challis},
  {Clocchiatti}, {Diercks}, {Filippenko}, {Gilliland}, {Hogan}, {Kirshner},
  {Leibundgut}, {Phillips}, {Reiss}, {Riess}, {Schmidt}, {Schommer}, {Smith},
  {Spyromilio}, {Stubbs}, {Suntzeff}, {Tonry}, \&
  {Carroll}}]{1998ApJ...509...74G}
{Garnavich}, P.~M., {Jha}, S., {Challis}, P., {Clocchiatti}, A., {Diercks}, A.,
  {Filippenko}, A.~V., {Gilliland}, R.~L., {Hogan}, C.~J., {Kirshner}, R.~P.,
  {Leibundgut}, B., {Phillips}, M.~M., {Reiss}, D., {Riess}, A.~G., {Schmidt},
  B.~P., {Schommer}, R.~A., {Smith}, R.~C., {Spyromilio}, J., {Stubbs}, C.,
  {Suntzeff}, N.~B., {Tonry}, J., \& {Carroll}, S.~M. 1998, \apj, 509, 74

\bibitem[{{Halverson} {et~al.}(2002){Halverson}, {Leitch}, {Pryke}, {Kovac},
  {Carlstrom}, {Holzapfel}, {Dragovan}, {Cartwright}, {Mason}, {Padin},
  {Pearson}, {Readhead}, \& {Shepherd}}]{2002ApJ...568...38H}
{Halverson}, N.~W., {Leitch}, E.~M., {Pryke}, C., {Kovac}, J., {Carlstrom},
  J.~E., {Holzapfel}, W.~L., {Dragovan}, M., {Cartwright}, J.~K., {Mason},
  B.~S., {Padin}, S., {Pearson}, T.~J., {Readhead}, A.~C.~S., \& {Shepherd},
  M.~C. 2002, \apj, 568, 38

\bibitem[{{Hanany} {et~al.}(2000){Hanany}, {Ade}, {Balbi}, {Bock}, {Borrill},
  {Boscaleri}, {de Bernardis}, {Ferreira}, {Hristov}, {Jaffe}, {Lange}, {Lee},
  {Mauskopf}, {Netterfield}, {Oh}, {Pascale}, {Rabii}, {Richards}, {Smoot},
  {Stompor}, {Winant}, \& {Wu}}]{2000ApJ...545L...5H}
{Hanany}, S., {Ade}, P., {Balbi}, A., {Bock}, J., {Borrill}, J., {Boscaleri},
  A., {de Bernardis}, P., {Ferreira}, P.~G., {Hristov}, V.~V., {Jaffe}, A.~H.,
  {Lange}, A.~E., {Lee}, A.~T., {Mauskopf}, P.~D., {Netterfield}, C.~B., {Oh},
  S., {Pascale}, E., {Rabii}, B., {Richards}, P.~L., {Smoot}, G.~F., {Stompor},
  R., {Winant}, C.~D., \& {Wu}, J.~H.~P. 2000, \apjl, 545, L5

\bibitem[{{Mason} {et~al.}(2003){Mason}, {Pearson}, {Readhead}, {Shepherd},
  {Sievers}, {Udomprasert}, {Cartwright}, {Farmer}, {Padin}, {Myers}, {Bond},
  {Contaldi}, {Pen}, {Prunet}, {Pogosyan}, {Carlstrom}, {Kovac}, {Leitch},
  {Pryke}, {Halverson}, {Holzapfel}, {Altamirano}, {Bronfman}, {Casassus},
  {May}, \& {Joy}}]{2003ApJ...591..540M}
{Mason}, B.~S., {Pearson}, T.~J., {Readhead}, A.~C.~S., {Shepherd}, M.~C.,
  {Sievers}, J., {Udomprasert}, P.~S., {Cartwright}, J.~K., {Farmer}, A.~J.,
  {Padin}, S., {Myers}, S.~T., {Bond}, J.~R., {Contaldi}, C.~R., {Pen}, U.,
  {Prunet}, S., {Pogosyan}, D., {Carlstrom}, J.~E., {Kovac}, J., {Leitch},
  E.~M., {Pryke}, C., {Halverson}, N.~W., {Holzapfel}, W.~L., {Altamirano}, P.,
  {Bronfman}, L., {Casassus}, S., {May}, J., \& {Joy}, M. 2003, \apj, 591, 540

\bibitem[{{Miller} {et~al.}(1999){Miller}, {Caldwell}, {Devlin}, {Dorwart},
  {Herbig}, {Nolta}, {Page}, {Puchalla}, {Torbet}, \&
  {Tran}}]{1999ApJ...524L...1M}
{Miller}, A.~D., {Caldwell}, R., {Devlin}, M.~J., {Dorwart}, W.~B., {Herbig},
  T., {Nolta}, M.~R., {Page}, L.~A., {Puchalla}, J., {Torbet}, E., \& {Tran},
  H.~T. 1999, \apjl, 524, L1

\bibitem[{{Ostriker} \& {Steinhardt}(1995)}]{1995Natur.377..600O}
{Ostriker}, J.~P. \& {Steinhardt}, P.~J. 1995, \nat, 377, 600

\bibitem[{{Peebles} \& {Ratra}(2003)}]{2003RvMP...75..559P}
{Peebles}, P.~J. \& {Ratra}, B. 2003, Reviews of Modern Physics, 75, 559

\bibitem[{{Perlmutter} {et~al.}(1999){Perlmutter}, {Aldering}, {Goldhaber},
  {Knop}, {Nugent}, {Castro}, {Deustua}, {Fabbro}, {Goobar}, {Groom}, {Hook},
  {Kim}, {Kim}, {Lee}, {Nunes}, {Pain}, {Pennypacker}, {Quimby}, {Lidman},
  {Ellis}, {Irwin}, {McMahon}, {Ruiz-Lapuente}, {Walton}, {Schaefer}, {Boyle},
  {Filippenko}, {Matheson}, {Fruchter}, {Panagia}, {Newberg}, {Couch}, \& {The
  Supernova Cosmology Project}}]{1999ApJ...517..565P}
{Perlmutter}, S., {Aldering}, G., {Goldhaber}, G., {Knop}, R.~A., {Nugent}, P.,
  {Castro}, P.~G., {Deustua}, S., {Fabbro}, S., {Goobar}, A., {Groom}, D.~E.,
  {Hook}, I.~M., {Kim}, A.~G., {Kim}, M.~Y., {Lee}, J.~C., {Nunes}, N.~J.,
  {Pain}, R., {Pennypacker}, C.~R., {Quimby}, R., {Lidman}, C., {Ellis}, R.~S.,
  {Irwin}, M., {McMahon}, R.~G., {Ruiz-Lapuente}, P., {Walton}, N., {Schaefer},
  B., {Boyle}, B.~J., {Filippenko}, A.~V., {Matheson}, T., {Fruchter}, A.~S.,
  {Panagia}, N., {Newberg}, H.~J.~M., {Couch}, W.~J., \& {The Supernova
  Cosmology Project}. 1999, \apj, 517, 565

\bibitem[{{Ratra} \& {Peebles}(1988)}]{1988PhRvD..37.3406R}
{Ratra}, B. \& {Peebles}, P.~J.~E. 1988, \prd, 37, 3406

\bibitem[{{Riess} {et~al.}(1998){Riess}, {Filippenko}, {Challis},
  {Clocchiatti}, {Diercks}, {Garnavich}, {Gilliland}, {Hogan}, {Jha},
  {Kirshner}, {Leibundgut}, {Phillips}, {Reiss}, {Schmidt}, {Schommer},
  {Smith}, {Spyromilio}, {Stubbs}, {Suntzeff}, \&
  {Tonry}}]{1998AJ....116.1009R}
{Riess}, A.~G., {Filippenko}, A.~V., {Challis}, P., {Clocchiatti}, A.,
  {Diercks}, A., {Garnavich}, P.~M., {Gilliland}, R.~L., {Hogan}, C.~J., {Jha},
  S., {Kirshner}, R.~P., {Leibundgut}, B., {Phillips}, M.~M., {Reiss}, D.,
  {Schmidt}, B.~P., {Schommer}, R.~A., {Smith}, R.~C., {Spyromilio}, J.,
  {Stubbs}, C., {Suntzeff}, N.~B., \& {Tonry}, J. 1998, \aj, 116, 1009

\bibitem[{{Riess} {et~al.}(2004){Riess}, {Strolger}, {Tonry}, {Casertano},
  {Ferguson}, {Mobasher}, {Challis}, {Filippenko}, {Jha}, {Li}, {Chornock},
  {Kirshner}, {Leibundgut}, {Dickinson}, {Livio}, {Giavalisco}, {Steidel},
  {Ben{\'{\i}}tez}, \& {Tsvetanov}}]{2004ApJ...607..665R}
{Riess}, A.~G., {Strolger}, L., {Tonry}, J., {Casertano}, S., {Ferguson},
  H.~C., {Mobasher}, B., {Challis}, P., {Filippenko}, A.~V., {Jha}, S., {Li},
  W., {Chornock}, R., {Kirshner}, R.~P., {Leibundgut}, B., {Dickinson}, M.,
  {Livio}, M., {Giavalisco}, M., {Steidel}, C.~C., {Ben{\'{\i}}tez}, T., \&
  {Tsvetanov}, Z. 2004, \apj, 607, 665

\bibitem[{{Spergel} \& {Pen}(1997)}]{1997ApJ...491L..67S}
{Spergel}, D. \& {Pen}, U. 1997, \apjl, 491, L67+

\bibitem[{{Spergel} {et~al.}(2003){Spergel}, {Verde}, {Peiris}, {Komatsu},
  {Nolta}, {Bennett}, {Halpern}, {Hinshaw}, {Jarosik}, {Kogut}, {Limon},
  {Meyer}, {Page}, {Tucker}, {Weiland}, {Wollack}, \&
  {Wright}}]{2003ApJS..148..175S}
{Spergel}, D.~N., {Verde}, L., {Peiris}, H.~V., {Komatsu}, E., {Nolta}, M.~R.,
  {Bennett}, C.~L., {Halpern}, M., {Hinshaw}, G., {Jarosik}, N., {Kogut}, A.,
  {Limon}, M., {Meyer}, S.~S., {Page}, L., {Tucker}, G.~S., {Weiland}, J.~L.,
  {Wollack}, E., \& {Wright}, E.~L. 2003, \apjs, 148, 175

\bibitem[{{Vilenkin} \& {Shellard}(1994)}]{1994csot.book.....V}
{Vilenkin}, A. \& {Shellard}, E.~P.~S. 1994, {Cosmic strings and other
  topological defects} (Cambridge Monographs on Mathematical Physics,
  Cambridge: Cambridge University Press, |c1994 ISBN 0521391539.)

\bibitem[{{Weinberg}(1989)}]{1989RvMP...61....1W}
{Weinberg}, S. 1989, Reviews of Modern Physics, 61, 1

\end{thebibliography}
\bibliographystyle{apj}

\end{document}